\documentclass[aps,prb,twocolumn,showpacs,superscriptaddress,groupedaddress]{revtex4}
\usepackage{graphicx}
\usepackage{fancyhdr}
\usepackage{amsmath}

\begin{document}

\title{Static and dynamic properties of vortex pairs in asymmetric nanomagnets} 

\author{B. C. Koop}
\affiliation{Royal Institute of Technology, 10691 Stockholm, Sweden}
\author{M. Gruschke}
\affiliation{Royal Institute of Technology, 10691 Stockholm, Sweden}
\author{T. Descamps}
\affiliation{Royal Institute of Technology, 10691 Stockholm, Sweden}
\author{A. Bondarenko}
\affiliation{Institute of Magnetism, National Academy of Science, 03142 Kiev, Ukraine}
\author{B. A. Ivanov}
\affiliation{Institute of Magnetism, National Academy of Science, 03142 Kiev, Ukraine}
\author{V. Korenivski}
\affiliation{Royal Institute of Technology, 10691 Stockholm, Sweden}

\date{\today}

\begin{abstract}
Stacked spin-vortex pairs in magnetic multilayered nanopillars, with vertical separation between the vortices small compared to the vortex core size, exhibit spin dynamics absent in individual vortices. This dynamics is nonlinear and is due to the strong direct core-core coupling in the system, dominating energetically for small-signal excitation. We observe and explain the appearance of spin resonance modes, forbidden within linear dynamics, and discuss how they depend on the magnetic and morphological asymmetries in the samples.
\end{abstract}


\maketitle 

\section{Introduction}
Spin vortices have been attracting a great deal of attention from the research community due to their high thermal stability, multiple stable configurations, well defined localized core-structure and high-quality resonance properties. A number of magnetic nanostructures have been investigated in detail with the aim to optimize and utilize the unique characteristics of magnetic vortices for use in, for example, spin-torque oscillators\cite{stno1,stno2,stno3} or magnetic recording\cite{switch1,switch2,switch3,switch4,switch5}. Most recently, systems of coupled vortices have gained much attention. This includes lateral arrays of magnetic particles with dipole-coupled vortex-cores\cite{array1,array2,array3,array4,array5} (far-field), vortex-pairs in a single magnetic particle\cite{double1,double2,double3,double4}, and vertically stacked vortices\cite{stack1,stack2,stack3,stack4}. All these structures have distinct features when it comes to their dynamic behavior, with the main resonant response typically being the gyroscopic resonance, in which the vortex-core moves in the particle plane in a trajectory large in radius compared to the size of the core. However, in the case of closely separated vertically stacked vortex-pairs, the direct core-core interaction can significantly alter the response to external fields. Whenever the separation of the two magnetic layers is significantly smaller than the size of the vortex cores, the core-core interaction at equilibrium is predominantly monopolar. We have previously shown that this coupling leads to high-frequency rotational and vibrational resonance modes \cite{PRL_109_097204},  in symmetric synthetic antiferromagnets.

\begin{figure}
\center
\includegraphics[width=1.\linewidth]{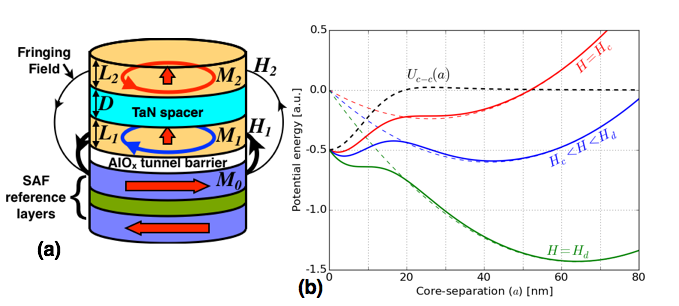}
\caption{(a) Schematic of the sample structure. (b) The total potential energy for identical layers as a function of the lateral core-separation (solid lines) for different external fields. $H_{c,dec}$ are the coupling and decoupling fields. The energy in the absence of the core-core interaction is displayed as colored dashed lines. The black dashed line displays the core-core interaction potential $U_\text{c-c}(a)$.\label{fig:schem_pot}}
\end{figure}

In this work we additionally consider asymmetric bi-layers, the so-called synthetic ferrimagnets, having a thickness-imbalance and/or an asymmetric bias-field applied. We compare analytic predictions with micromagnetic simulations and experimental results. The micromagnetic package used is MicroMagnum \cite{micromagnum}. For all simulations the samples were 420x350 nm elliptical disks with varying thicknesses. The cell-sizes used were  $\{x,y,z\} = \{2.0,2.0,2.5\}$ nm. The measurements were done on samples as schematically depicted in figure \ref{fig:schem_pot}a. The quantity measured on the experiment is the DC resistance ($Z$) of a synthetic nanomagnetic tunnel junction: $Z\propto\langle\boldsymbol{M_0}\rangle\cdot\langle\boldsymbol{M_1}\rangle\propto\langle M_{1x}\rangle$. The lateral dimensions of our samples are 420x350 and 450x375 nm. The layer-thicknesses are in the range of {$L_1,D,L_2$} = {5, 1, 5} nm.

\section{Vortex-pair quasi-statics}
Synthetic antiferromagnets (SAF) and SAF's with geometric or magnetic asymmetry, also known as synthetic ferrimagnets (SFi), can significantly differ in static and dynamic behavior. We have previously demonstrated this for the antiparallel (AP) ground state of the SFi, in which the individual particles are magnetized uniformly \cite{APL_asym,IEEE_asym}. Here we discuss the influence of asymmetry on the quasistatic behavior of the vortex-pair states found in SFi and, specifically, the states with parallel cores and antiparallel chiralities (P-AP).
The field asymmetry we consider is a difference in the in-plane DC-field acting on the two layers in a SAF. This asymmetry can originate from a not fully flux-closed read-out junction, shown schematically in Fig. \ref{fig:schem_pot}a.

\begin{figure*}
\includegraphics[width=0.95\linewidth]{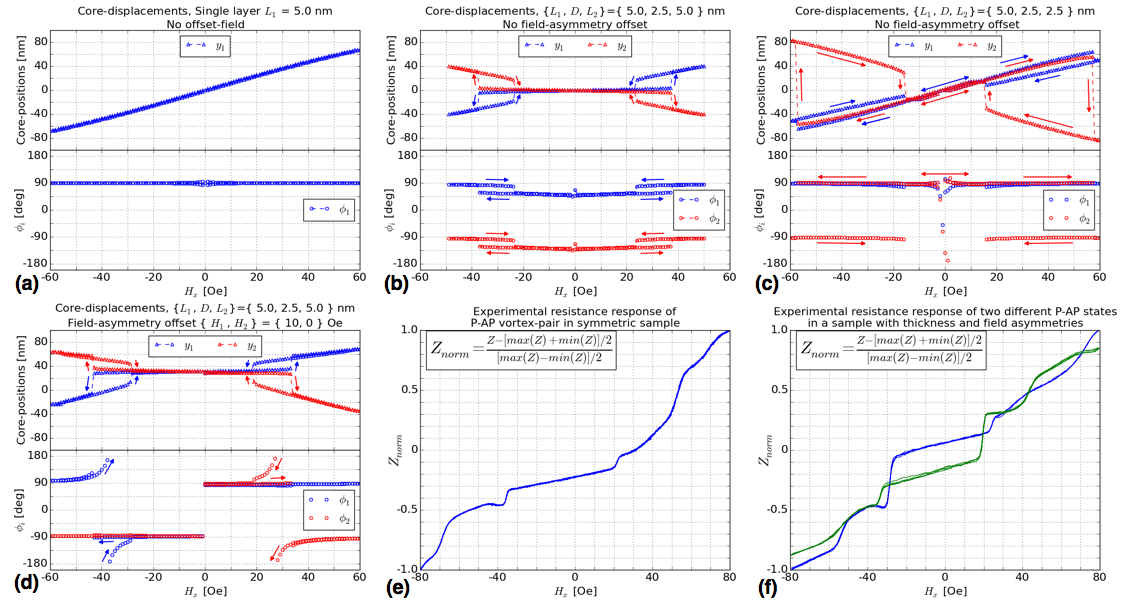}
\caption{(a-d) Micromagnetically simulated core-displacement versus applied DC-field ($H_x$) for (a) single layer, (b) SAF, (c) thickness-asymmetric SFi, and (d) bias-field-asymmetric SFi. The top panels in (a-d) show the projection of the core-displacement on the Y-axis, while the bottom panels display the angle of the displacement with respect to the applied field. (e) and (f) show the measured P-AP state resistance vs field for (e) an almost symmetric sample (SAF) and (f) a sample with clear thickness and field asymmetries (SFi).\label{fig:DC_complete}}
\end{figure*}

In the absence of an external field the vortex core in a single ferromagnetic particle, typically thin and circular or elliptical in-plane, is found in the center of the particle due to the interaction of the vortex with the particle boundary. An external in-plane field results in a core displacement off the particle center, perpendicular to the direction of the applied field. This has been demonstrated for small core displacements ($r/R \ll 1$) in circular particles \cite{guslienko}. For elliptical particles the direction of the displacement is perpendicular to the field if the field is applied along the long axis of the ellipse ($R_a$) - the case of interest here. The core displacement is then along the short axis of the ellipse \cite{PRB_74_064404} ($R_b$), as shown in Fig. \ref{fig:DC_complete}a). The energy is given by
\begin{equation}
U_{i}(y_{i})=U_0+\frac{20}{9}\pi M_S^2L_{i}^2y_{i}^2/R_b -\pi M_S L_{i} H_xR_a \sigma y_{i}, \label{eq:U_single}
\end{equation}
where $i=1,2$ denotes an individual layer. $U_0$ is a contribution essentially independent on the core-position, $R_{a,b}$ the long and short axes radii of the ellipse, $L$ the thickness of the disk, $M_S$ the saturation magnetization, $H_x$ the external field applied along the long axis (x-axis) of the sample, and $\sigma=\pm1$ the chirality of the vortex (clockwise = -1, counterclockwise = +1).

When considering two stacked vortices in close proximity ($L_{1,2}/D\ll 1$), with negligible interlayer exchange coupling (our case for the TaN spacer), the cores experience predominantly monopolar interaction when the inter-core lateral displacement is smaller than the core-diameter\cite{PRL_109_097204}. In this work we focus on the P-AP vortex-pair state, with parallel core orientations and  antiparallel chirality. Due to the opposite chirality, an applied easy-axis field acts to displace the cores into opposite directions. In the weak-field limit, this displacement is suppressed by the strong quasi-monopolar coupling between the two cores aligned in parallel, i.e., when the Zeeman energy for individual cores is smaller than the core-core interaction energy ($U_\text{c-c}(a)$, derived in \cite{PRL_109_097204}, with $a$ being the lateral core-core separation). 

Figure \ref{fig:schem_pot}b shows the potential energy of the core-core coupling as a function of the core-separation (black dashed line). This potential energy is independent of the external field. The total potential, including the Zeeman and the core-core terms for identical layers ($a=y_i/2$), $U_\text{tot}(a) = U_1(a)+U_2(a)+U_\text{c-c}(a)$, is depicted in figure \ref{fig:schem_pot}b as solid lines for different values of the applied field. Two characteristic fields, $H_\text{c}$ and $H_\text{dec}$, are the coupling and decoupling fields, respectively, and correspond to the transition points between one and two minima in the potential (saddle points). Since $H_\text{c}<H_\text{dec}$, there is a field range with both states stable or meta-stable, which naturally gives rise to hysteresis.

We use micromagnetic simulations to characterize the vortex-core movement and find the existence of hysteresis for SAF particles. Figure \ref{fig:DC_complete}b shows the existence of well-defined bistable states in the region of $23 < |H_x| < 38$ Oe. Starting from zero applied field, the displacement of a core in a coupled core-pair is linear in field, with a slope of the order of 0.1 nm/G. When the field reaches the threshold value at which the energy minimum for the core-pair becomes a saddle point ($H_x=H_{dec}$), the two cores decouple and move to their respective equilibrium position dictated by the Zeeman energy for individual vortices. The core-core interaction is dipolar in nature at separations much larger than the core size. Once decoupled, the cores remain far apart until the field is reduced passed the threshold for the Zeeman-minimum to become a saddle point ($H_x=H_c$). In agreement with our analytic prediction, the micromagnetic analysis shows that there is a field region where both the coupled and decoupled state of the two cores in the SAF system are stable. The actual core-core state in this field region is determined by the magnetic history of the sample.

Experimentally, for symmetric SAF samples we observe a linear response in the junction resistance (linear core displacement) for small fields, followed by a step corresponding to core-core decoupling, as shown in figure \ref{fig:DC_complete}e. The decoupling is relatively field-symmetric (apart from a small offset). However, the coupling-decoupling transition in resistance is not as abrupt as observed micromagnetically. Additionally, the hysteresis behavior is not present. We note that the experiments are performed at room temperature. When the potential barrier to core-decoupling at a given applied field becomes small (for $H_x\sim \{H_c,H_{dec}\}$), thermal agitation can activate transitions between the two minima in the system. This should produce effective averaging in the measured junction resistance ($Z_{dc}$): $Z_{dc} = \int_{0}^{T}{Z(t)dt}/T= Z_{c}P_{dec}+Z_{dec}P_{c}$, where $P_{c,dec}$ is the probability of coupling/decoupling, $Z_{c,dec}$ - the coupled/decoupled resistance.

Introducing a thickness-imbalance between the two particles in a SAF results in significantly altered trajectories for uncoupled cores as a function of the applied field. As can be seen from eq. \ref{eq:U_single}, the individual-core displacement increases with decreasing the thickness of the layer for a given field, effectively increasing the core susceptibility (due to the smaller core volume). In the coupled core system, this results in dragging of the 'smaller' core (thin layer) by the 'bigger' core (thick layer). Figure \ref{fig:DC_complete}c shows the simulated core displacements for thicknesses $L_1 = 5$ nm and $L_2 = 2.5$ nm. As expected, the two cores initially move together as a pair. The thin-core follows the thick-core resulting in a significant increase of the field-coefficient of the core displacements.

The decrease in the demagnetizing and Zeeman energies of the thin layer, acting to break the core-core pair in the P-AP state, results in an increase of the decoupling field ($H_{dec}$). After decoupling, the cores are further apart from each other compared to the case of SAF, the core-core potential is reduced, and the field of recoupling ($H_{c}$) is therefore reduced, giving an increased field-range of bistability (larger hysteresis).
Interestingly, the displacement of a decoupled core in the thick layer is smaller than that of a coupled core for a given field $H_x$. This non-trivial effect is due to the interplay between the core-core and core-boundary interactions, specific to the asymmetric SFi system.

Unlike thickness-asymmetry, bias-field asymmetry results in a field offset in the core displacement, such that $y_1(H_x) \neq -y_1(-H_x)$. This is confirmed by our simulation results shown in figure \ref{fig:DC_complete}d, where an additional small fixed biasing field was applied to the bottom layer ($H_{1x}$). At zero-field the two cores are already displaced off the particle center, however their individual field-coefficients are the same as in the symmetric case. Due to the initial core offset, the field response of a coupled pair is slightly shifted, which can be seen as asymmetric displacement in the decoupled region for positive and negative applied fields. Asymmetry in biasing thus results in both offset positions and offset separations of the cores.

Measurements on a sample having a finite asymmetric biasing field from the readout junction as well as thickness asymmetry show a similar core displacement behavior as that obtained numerically for SFi (fig. \ref{fig:DC_complete}f). The two curves in figure \ref{fig:DC_complete}f are the resistance response of two different P-AP states in the same sample. The difference in the coupling/decoupling step height for opposite fields agrees with the numerical response for the case of bias-field asymmetry. However, a bias-field alone can not lift the 4-fold degeneracy of the P-AP state. Therefore, the difference in the field response seen in the measured data for two P-AP states implies the existence of a small thickness imbalance lifting the degeneracy.

\begin{figure*}
\includegraphics[width=0.7\linewidth]{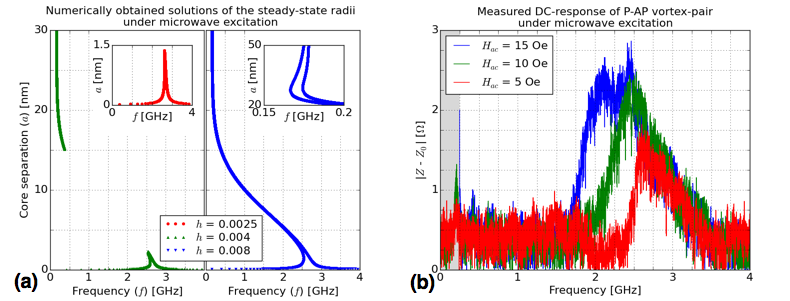}
\caption{(a) Amplitude of the steady-state gyration for different excitation amplitudes obtained from equation \ref{eq:res_freq}. (b) Measured spectroscopic response for different excitation amplitudes. The frequency is swept from 4 GHz down to 1 MHz. The gray area denotes the frequency range for $H_{ac}=15$ Oe not included in the plot due to vortex annihilation.\label{fig:AC_complete}}
\end{figure*}

\section{Vortex-pair dynamics}
In our case of strongly coupled vortex-cores (core lengths of 5 nm and core-core spacing of 1 nm, on axis, with no direct exchange via the 1 nm thick TaN spacer), the dynamic response to an alternating in-plane magnetic field is of similar character as that of the well-known gyroscopic mode for single vortices \cite{} and in-plane vortex-pairs \cite{}, albeit with greatly different energetics. Indeed, due to the strong quasi-monopole core-core coupling the frequency of the rotational resonance, where the two cores are coupled and gyrate about each other, is an order of magnitude higher than the gyroscopic frequency for an individual (decoupled) core. We have previously analyzed the low-amplitude limit of this coupled-core rotational mode \cite{PRL_109_097204} and reported preliminary experimental evidence of non-linearities in the rotational resonance \cite{IEEE_vortex}. 

Here we study in more detail the nonlinear dynamics of a near-symmetric P-AP vortex-pair. Due to the fact that the coupling is strongly localized (see Fig. 1b), the actual resonance frequency decreases with core-core separation. This means that, if the excitation amplitude is tuned carefully, every frequency between the traditional gyroscopic frequency ($\sim 200$ MHz) and the zero-amplitude rotational frequency ($\sim 2.5$ GHz) can have a stationary solution. Our analytic derivation to be reported elsewhere yields the following expression for the stationary trajectory amplitude for different frequencies and amplitudes:
\begin{equation}
 f = \frac{\omega(a)}{1+\lambda^2}\left[1\pm\sqrt{1-(1+\lambda^2)\left(1-\frac{h^2}{\omega^2 (a)a^2}\right)}\right],\label{eq:res_freq}
\end{equation}
where $\omega (a) $ is the frequency (normalized by $\gamma M_s$) of non-linear gyration of the cores with equal separations $a$ without damping and ac-field;
$\lambda-$ the dissipation constant; $h=2RH_{ac}/(4\pi M_s \Delta) -$ the normalized field amplitude with $M_s$ being the saturation magnetization, $R$ the disk-radius, $\Delta$ the core-radius, and $H_{ac}$ the microwave field amplitude.

Numerically solving equation \ref{eq:res_freq} yields for low excitation amplitudes a slight asymmetric tilt of the rotational peak ($f_{rot}$) toward lower frequencies. The core-core separation $a$ of the stationary trajectories is in the range of 0 to 3 nm, which is in the range of the strong quasi-monopolar inter-core coupling. Increasing the excitation amplitude (Fig. \ref{fig:AC_complete}a, $h=0.04$), increases the tilt of the resonant peak around $f_{rot}$, resulting in an effective decrease of the peak-frequency. Additionally, at low-frequencies, around the characteristic gyroscopic frequency, large-radius trajectories become stable solutions. These low-frequency resonances can be excited and stabilized if the cores can experience, for example, occasional stochastic decoupling. The low-field excitation by itself is insufficient to decouple the cores. However, at finite temperature, decoupling may occur via thermal agitation since the magnetic volume of the vortex core is small (non-vanishing $kT$ compared to $U_\text{c-c}$). Experimentally, at room temperature, we do observe excitation of the gyroscopic mode at low field amplitudes from the nominally coupled core-core  state (Fig. \ref{fig:AC_complete}b, $H_{ac}=10$ Oe; green peak at $\sim 200$ MHz).

When the AC-amplitude is increased further (Fig. \ref{fig:AC_complete}a, $h = 0.08$), the model predicts that the two resonant areas become connected and that there is a bi-stable state for all frequencies between $f_{gyr}$ and $f_{rot}$. Experimentally, the response for decoupled cores can only be obtained for relatively small amplitudes since for large excitation amplitudes the vortex-trajectories are too large and result in vortex annihilation. This is seen in Fig. \ref{fig:AC_complete}b for an excitation amplitude of 15 Oe. Here the vortex was annihilated while excited at 250 MHz, which is slightly above the gyroscopic resonance. Increasing the excitation amplitude further results in a broadening of the annihilation region toward higher frequencies.

\section{Conclusions}
The effects of thickness and bias-field asymmetry on spin vortex-pair properties in a SAF particle are investigated. We focus on the parallel-core/antiparallel-chirality state, where the core-core coupling is strong and can dominate energetically. We show both theoretically and experimentally how the presence of asymmetry lifts the degeneracy of different P-AP states. Numerically we find a bi-stable state, where both coupled and decoupled core states are stable for a given DC-field range. The resulting hysteretic behavior is absent in our experimental results due, presumably, to thermal activation of the core-core coupling-decoupling transitions.

We analytically predict and experimentally confirm nonlinear dynamic behavior of the main rotational resonance, even in the weak-excitation limit. We find stable large-amplitude steady-state trajectories, which can be excited via thermally activated decoupling of a core-core pair. Experimentally, this is observed as low-frequency resonance peaks in the nominally strongly coupled core-core regime for a spin-vortex pair in a synthetic nanomagnet.

\subsection*{acknowledgment}
This research was funded by the Swedish Research Council, project number 2014-4548.

\bibliography{aiptemplate}

\end{document}